\documentclass[twoside]{article}
\usepackage{fleqn,espcrc2,epsfig}

\usepackage{graphicx}
\usepackage[figuresright]{rotating}

\newcommand{\ba}{\begin{eqnarray}}
\newcommand{\ea}{\end{eqnarray}}
\newcommand{\be}{\begin{equation}}
\newcommand{\ee}{\end{equation}}
\newcommand{\bi}{\begin{itemize}}
\newcommand{\ei}{\end{itemize}}

\newcommand{\middle}[1]{\raisebox{1.5ex}[0pt]{#1}}

\newcommand{\fig}[1]{figure \ref{fig:#1}}

\newcommand{\Tab}[1]{Table \ref{tab:#1}}

\newcommand{\ZV}{Z_V^{\rm eff}}

\font\tenmsb=msbm10 scaled\magstep1
\font\sevenmsb=msbm7 scaled\magstep1
\font\fivemsb=msbm5 scaled\magstep1
\newfam\msbfam
\textfont\msbfam=\tenmsb
\scriptfont\msbfam=\sevenmsb
\scriptscriptfont\msbfam=\fivemsb

\newcommand{\order}[1]{{\mathcal O}(#1)}

\newcommand{\vk}{v\cdot k}

\newcommand{\vub}{V_{\rm ub}}

\title{\hfill\begin{minipage}{0pt}\scriptsize \begin{tabbing}
	\hspace*{\fill} Edinburgh-1999/12\\ \end{tabbing}\end{minipage}\\[8pt]
	\vspace{-0.8cm}
	Determination of $\vub$ from $B\to\pi l\bar{\nu}$ on the lattice}

\author{UKQCD Collaboration\\
	Presented by C.M.~Maynard\address{Department of Physics and Astronomy,
	University of Edinburgh EH9 3JZ, UK}}
       
\begin{document}

\begin{abstract}
We present results of a lattice study of the form factors in the decay 
$B\to \pi l\bar{\nu}$. We attempt to disentangle the the dependence of the form
factors on the light quark masses and the momentum transfer. Using models of 
the $q^2$ dependence we calculate the total decay rate, and compare to the
experimental measure to extract $\vub$. This study was performed in the
quenched approximation at $\beta=6.2$ on a $24^3\times 48$ lattice, with a
non-perturbatively improved SW fermion action.
\vspace{1pc}
\end{abstract}

\maketitle

\section{INTRODUCTION}
Semileptonic decays of mesons containing a $b$ quark play an important role
in the determination of Cabibbo-Kobayashi-Maskawa (CKM) matrix elements. 
The transition amplitude of the decay $B\to\pi l\bar{\nu}$ factorizes into 
leptonic and hadronic parts. This hadronic matrix element can be parameterised
by two form factors
\ba
  \langle \pi(\vec{k})|V^{\mu}|B(\vec{p})\rangle 
	&=&f_+(q^2)(p+k-q\Delta_{m^2})^{\mu}	\nonumber \\
   &+&f_0(q^2)q^{\mu}\Delta_{m^2}
\ea
where $\Delta_{m^2}=(m_B^2-m_{\pi}^2)/q^2$ and $q=p-k$. In the limit of zero
lepton mass, the total decay rate is given by
\be
\label{eqn:tot_decay}
\Gamma=\frac{G_F^2|\vub|^2}{192\pi^3m_B^3}
	\int^{\eta^2}_0 [\lambda(q^2)]^{3/2}|f_+(q^2)|^2 dq^2
\ee
where $\eta^2=(m_B-m_{\pi})^2$ and
\be
  \lambda(q^2)=(m_B^2+m_{\pi}^2-q^2)^2-4m_B^2m_{\pi}^2.
\ee
We can determine the decay rate from the $q^2$ dependence of the form factor
$f_+(q^2)$ and then compare to the experimental measure of the decay 
rate~\cite{CLEO_B2pi} to extract $\vub$.

\section{DETAILS OF THE CALCULATION}
The 216 gauge quenched configurations were generated using the Wilson action
on a $24^3\times 48$ lattice. 
The quark propagators were calculated using an $\order{a}$ improved
action, where the coefficient $c_{SW}$ has been determined 
non-perturbatively~\cite{alpha_np} (NP). We use four heavy quarks with masses
around charm, $(\kappa_H=0.1200,0.1233,0.1266,0.1299)$. Three light quarks
with masses around strange $(\kappa_L=0.1346,0.1351,0.1353)$ are used
for the active propagator, and the heaviest two for the spectator. The heavy
quarks were smeared~\cite{boyling_p} and the light quarks fuzzed. The chiral 
limit has been determined to be~\cite{ukqcd_prep} $\kappa_{\rm crit}=0.135815$,
and the physical value of $m_{\pi}/m_{\rho}$ corresponds to $\kappa_n=0.13577$.
The lattice spacing is set by $m_{\rho}$ and $a^{-1}=2.64$ GeV.

We obtain the form factors from the heavy-to-light three-point correlation
functions, using the masses and amplitudes from the heavy-light and light-light
two-point correlation functions. The general method is given 
in~\cite{bowler_d2k}. We place the operator for the heavy-light pseudoscalar 
meson at $T=20$ rather than the mid-point of the lattice to check for 
contamination from different time orderings. We use eight different 
combinations of $\vec{p}$ and $\vec{k}$ determine the $q^2$ dependence; 
$0\to0$, $0\to1$, $0\to\sqrt 2$, $1\to0$, $1\to1$, $1\to1_{\bot}$, $
1\to1_{\gets}$ and $1\to\sqrt 2_{\bot}$ in lattice units. There is no $0\to0$ 
channel for $f_+$.

\subsection{Mass dependent renormalisation}
We can also remove all $\order{a}$ errors from matrix elements of on-shell 
states by an appropriate definition of the currents. For the vector
current for degenerate quarks of mass $m_Q$, we have 
\begin{equation}
  V^R_{\mu} = Z_V(1+b_V am_Q)\{V_{\mu}+c_Va 
        \frac{1}{2}({\partial}_{\nu}+{\partial}_{\nu}^{\star})T_{\mu\nu}\}
\end{equation}
where $V_{\mu}$ and  $T_{\mu\nu}$ are the local lattice vector and tensor 
currents respectively. Both $b_V$ and $Z_V$ have
been determined non-perturbatively~\cite{alpha_np3_Z}. Preliminary results for
a non-perturbative determination of the mixing coefficient $c_V$ exist, but 
we use the one-loop perturbative estimate
, which is small. Defining
\begin{equation}
\label{eqn:ZV_eff}
  Z_V^{\rm eff}\equiv Z_V(1+b_V am_Q).
\end{equation}
For the forward degenerate matrix element, we can calculate $\ZV$ from our 
data. We show the comparison to $\ZV$ evaluated for these quark masses 
using the non-perturbative $Z_V$ and $b_V$ in \Tab{zv_eff}. The excellent
agreement suggests higher order discretisation effects are limited.
\begin{table}
\begin{center}
\label{tab:zv_eff}
\caption{The effective matching coefficient. The current does not depend on 
	$\kappa_S$.} 
\begin{tabular}{cccc}
      & \multicolumn{3}{c}{$Z_V^{\rm eff}$} \\\cline{2-4}
\middle{$am_Q$}& $\kappa_{S1}$ & $\kappa_{S2}$ & $Z_V(1+b_Vam_Q)$ 
	\\ \hline
$0.4852$& $1.316^{+3}_{-3}$& $1.317^{+4}_{-5}$ &$1.335$\\ \hline
$0.2680$& $1.093^{+2}_{-5}$&$1.087^{+2}_{-2}$&$1.093$
\vspace{-1.0cm}
\end{tabular} 
\end{center}
\end{table}

\section{CHIRAL EXTRAPOLATION}
 To evaluate the form factor $f_i$ at 
physical quark masses we must consider both the intrinsic dependence of $f_i$
and the indirect mass dependence arising from the change in $q^2$:
\begin{equation}
f_i=f_i(q^2_{(\kappa_A,\kappa_S)},\kappa_A,\kappa_S).
\end{equation}
In previous UKQCD analyses ~\cite{chrism} the $q^2$ dependence was modelled by
an extra term. This is potentially difficult to control. Here we extrapolate
whilst holding $q^2$ fixed. This approach yields a more reliable extrapolation.
This is discussed in more detail in ~\cite{vil}.

We first {\itshape interpolate} the form factors to a chosen set of $q^2$ 
values for each quark mass combination. The values of $q^2$ are chosen such 
that we interpolate for each light quark combination and that for different 
heavy quark masses, the sets of $q^2$ values correspond to the heavy quarks 
having the same {\itshape velocity}. This is discussed in the section on the
heavy extrapolation.

\begin{figure}[ht!]
\begin{center}
\vspace{-0.5cm}
\epsfig{figure=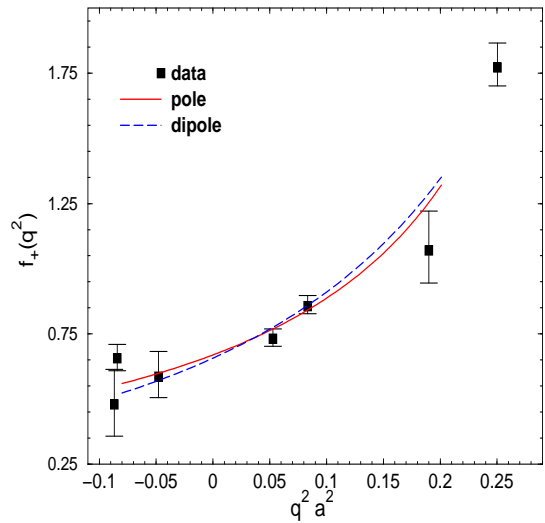,height=7cm,width=7cm,angle=-90} 
\vspace{-1.0cm}
\caption{\itshape $f_+(q^2)$ for $\kappa_H=0.1233$, $\kappa_A=0.1353$ and 
	$\kappa_S=0.1351$.  The support of the curves shows the 
	range $q^2$.}
\label{fig:fix_mom}
\vspace{-1.0cm}
\end{center}
\end{figure}

The form of the interpolation function is motivated by pole dominance models,
\be
\label{eqn:pole_dominance}
   f_i(q^2)=\frac{f_i(0)}{(1-q^2/m_i^2)^{n_i}}.
\ee
where $i$ is either $+$ or $0$.
However, as we interpolate in $q^2$, any model dependence in the chiral
extrapolation is mild, this is shown in figure \ref{fig:fix_mom}.
We then extrapolate the form factors at fixed $q^2$ to $\kappa_n$
with the light quarks non-degenerate;
\ba
\label{eqn:chiral_FF_k}
  f(\kappa_{S}, \kappa_{A})&=&\alpha + \beta \left(
  \frac{1}{\kappa_S}-\frac{1}{\kappa_{\rm crit}} \right )\nonumber \\
  &&+\gamma\left(\frac{1}{\kappa_S}+\frac{1}{\kappa_A}-
	\frac{2}{\kappa_{\rm crit}}\right ).
\ea

\section{HEAVY QUARK MASS EXTRAPOLATION}
Heavy quark effective theory (HQET) is used to motivate the form of the 
extrapolation to the $B$ meson scale. The scaling relations, $f_+\sim 
\sqrt{M}$ and $f_0\sim 1/\sqrt{M}$ are determined at fixed four-velocity, $v$.
Defining the recoil variable,
\begin{equation}
  v\cdot k= \frac{M_P^2 + m_{\pi}^2 - q^2}{2 M_P}
\end{equation}
we can then extrapolate the form factors at fixed $\vk$ to the $B$ meson scale:
\begin{equation}
\label{eqn:heavy_extrap}
  C f_i(v\cdot k)M_P^{s_i/2}=
        \gamma_i\left (1 +\frac{\delta_i}{M_P}
         + \frac{\epsilon_i}{M^2_P} \right)\
\end{equation}
where $s_i=-1$ when $i=+$, and $s_i=+1$ when $i=0$. The coefficient $C$ is
the logarithmic matching factor,
\begin{equation}
\label{eqn:HQET_match}
  C(M_P,m_B)=\left (\frac{\alpha_s(m_B)}{\alpha_s(M_P)}\right)
        ^{2/\beta_0}
\end{equation}
and $\beta_0=11$ in quenched QCD.

\section{RESULTS}
\begin{figure}[ht!]
\begin{center}
\vspace{-0.7cm}
\epsfig{figure=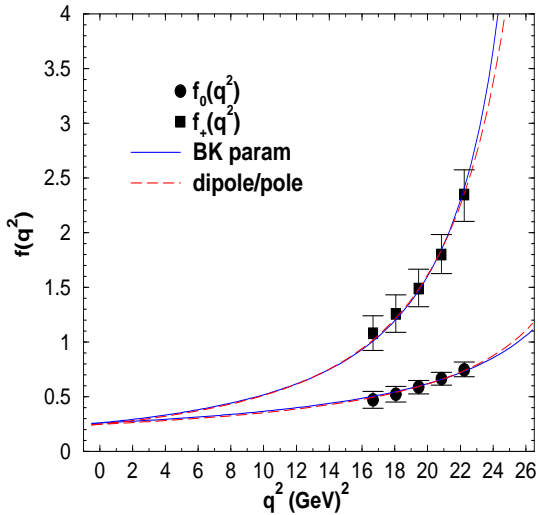,height=7cm,width=7cm} 
\vspace{-1.0cm}
\caption{\itshape The form factors for $B\to\pi l\nu$.}
\label{fig:B2pi_mom}
\vspace{-1.2cm}
\end{center}
\end{figure}

The resulting form factors are plotted in \fig{B2pi_mom}.
Pole dominance models, equation \ref{eqn:pole_dominance}, combined with the 
heavy quark scaling relations suggest that $n_+=n_0+1$. Light-cone scaling
further suggests $n_0=1$. We also impose the kinematic constraint $f_0(0)=
f_+(0)$, to parameterise the form factors by a pole for $f_0$ and a dipole
for $f_+$. A slightly more sophisticated
pole/dipole parameterisation for $f_0$ and $f_+$, consistent with the
same constraints, has been suggested by Becirevic and Kaidalov
(BK)~\cite{BK_param}:
\begin{eqnarray}
  f_+(q^2)&=&\frac{c_B(1-\alpha)}
  {(1-q^2/m^2_{B^{\star}})(1-\alpha q^2/m^2_{B^{\star}})} \nonumber \\
  f_0(q^2)&=&\frac{c_B(1-\alpha)}{(1- q^2/\beta m^2_{B^{\star}})}.      
\end{eqnarray}
We fit both parameterisations to the form factors. This is shown in figure
\ref{fig:B2pi_mom}. We can now use these models to calculate the total decay 
rate from equation \ref{eqn:tot_decay}. The results are
\be
\Gamma(B\to \pi l\nu)/|\vub|^2=9.0\pm3.0\pm3.2 ({\rm ps})^{-1}
\ee
where the first error is statistical and the second is systematic. The 
systematic errors are estimated by trying different interpolation functions
for the chiral extrapolation, a linear fit to the heaviest three quarks for
the heavy extrapolation and estimates of the lattice spacing from different
quantities, i.e. $r_0$.
We
can use this model dependent result to extract $\vub$ from experimental
data~\cite{CLEO_B2pi},
\be
  |\vub|=(3.7\pm0.5\pm0.7\pm0.7)\times 10^{-3}.
\ee
The third error is the experimental error in the branching ratio.

This is a preliminary model dependent result. The form factors are well
determined in the range $16-22\ {\rm Gev}^2$. The total decay rate is
dominated by low $q^2$ due to phase space. Here we have no data and are
reliant on models of $q^2$. The differential decay rate could be used to 
extract $\vub$ in a model independent manner, but there is no experimental
data available yet. This work was supported by EPSRC grant GR/K41663 and
PPARC grant GR/L29927.

\end{document}